\documentclass[%
reprint,
 amsmath,amssymb,
 aps,
prl,
]{revtex4-2}

\usepackage{graphicx}
\usepackage{dcolumn}
\usepackage{bm}
\usepackage{mathrsfs}
\usepackage{braket}
\usepackage{color}
\usepackage{comment}
\usepackage{siunitx}

\usepackage{hyperref}
\hypersetup{
     colorlinks   = true,
     linkcolor    = blue,
     citecolor    = blue,
     urlcolor     = blue
}

\newcommand{\up}{\uparrow}
\newcommand{\down}{\downarrow}

\begin{document}

\preprint{APS/123-QED}

\title{Thermodynamic relations for the Cooper pair's momentum in helical superconductors}

\author{Koki Shinada}
\altaffiliation[shinada.koki.64w@st.kyoto-u.ac.jp]{}
\author{Robert Peters}
\affiliation{
Department of Physics, Kyoto University, Kyoto 606-8502, Japan}%

\date{\today}

\begin{abstract}
Two thermodynamic relations are proposed for the exact measurement of the center-of-mass (COM) momentum of the Cooper pairs in helical superconductors. The first relation concerns the constraint on the change in the COM momentum in response to the variation in the magnetic field, which is linked to the superconducting Edelstein effect. The second relation is related to a first-order phase transition, showing that the jump of the COM momentum can be measured from the slope of the transition line on the phase diagram, where the supercurrent is used as a control parameter. These relations solve the difficulty of detecting the momentum, enabling broader and more precise exploration of noncentrosymmetric superconductors.
\end{abstract}

\maketitle

\textit{Introduction.}---
The emergence of novel superconducting phases is becoming increasingly diverse, and there is growing experimental and theoretical interest in superconductivity that extends beyond the framework of BCS theory. The diversification of superconducting gap symmetries beyond the $s$-wave is not the only focus; recently, attention has also been drawn to the finite center-of-mass (COM) momentum of Cooper pairs, which accompanies the spatial modulation of the superconducting gap in what is known as the Fulde-Ferrell-Larkin-Ovchinnikov (FFLO) state \cite{PhysRev.135.A550,LarkinOvchinnikov1964}. This FFLO state can be broadly classified into two types. One is the LO state, where the amplitude of the superconducting gap undergoes spatial modulation, behaving as $\Delta(\bm{r}) = \Delta \cos (\bm{q}\cdot \bm{r})$. This has been discussed with the original mechanism by the Zeeman magnetic field in various systems \cite{doi:10.1073/pnas.1413477111,PhysRevLett.124.107001,doi:10.1143/JPSJ.76.051005,https://doi.org/10.1002/andp.201700282}. The other is the FF state, which involves a phase modulation that behaves as $\Delta(\bm{r}) = \Delta e^{i\bm{q} \cdot \bm{r}}$. The FF state has been actively studied, particularly in noncentrosymmetric superconductors with antisymmetric spin-orbit coupling (SOC), and it is especially known as helical superconductor \cite{AGTERBERG200313,dimitrova2003phase,PhysRevLett.94.137002,PhysRevB.75.064511,PhysRevB.76.014522,BauerSigrist2012}. 
Several candidate materials for the helical superconductivity have been identified, including monolayer Pb \cite{PhysRevLett.111.057005}, doped $\mathrm{SrTiO_3}$ \cite{PhysRevB.101.100503}, artificial superlattices \cite{PhysRevB.96.174512,ando2020observation,narita2022field,Kawarazaki_2022}, van der Waals heterostructure \cite{PhysRevResearch.5.L022064}, and twisted graphene \cite{lin2022zero}. 

Despite the growing number of candidate materials for helical superconductivity, research on its detection remains in its infancy.
In these systems, the robustness against in-plane magnetic fields has been observed, and the consistency of their phase diagrams with theoretical predictions provides indirect evidence of a helical superconductor \cite{PhysRevB.101.100503,PhysRevB.96.174512}. 
Additionally, there has been growing interest in studying transport phenomena, with ongoing efforts to achieve a more direct detection. 
In the superconducting diode effect \cite{ando2020observation,PhysRevLett.128.037001,doi:10.1073/pnas.2119548119,He_2022,narita2022field,Kawarazaki_2022,PhysRevResearch.5.L022064,lin2022zero}, for instance, the sudden reversal of sign due to a sharp increase in the COM momentum under high magnetic fields has been discussed. Furthermore, a recent measurement of nonlinear resistivity has observed an enhancement at high magnetic fields \cite{asaba2024evidence}.
These findings verify the momentum crossover, a characteristic feature of helical superconductivity, bringing us closer to confirming its existence. However, if we seek more direct evidence, we must return to the fundamental definition: the direct detection of the finite COM momentum is essential. Nevertheless, since the helical superconducting state is the FF state, where the modulation pertains to the phase of the superconducting gap, direct observation of this state is not straightforward, except for a few proposals that utilize interference effects \cite{PhysRevLett.84.4970,PhysRevLett.94.137002,Chen_2014}.

In this letter, to overcome the difficulties of its detection, we propose two thermodynamic relations to exactly measure the COM momentum in all helical superconductors. Furthermore, these relations are specifically studied in a two-dimensional Rashba superconductors.

\textit{Model.}---
\begin{figure*}[t]
\includegraphics[width=0.31\linewidth]{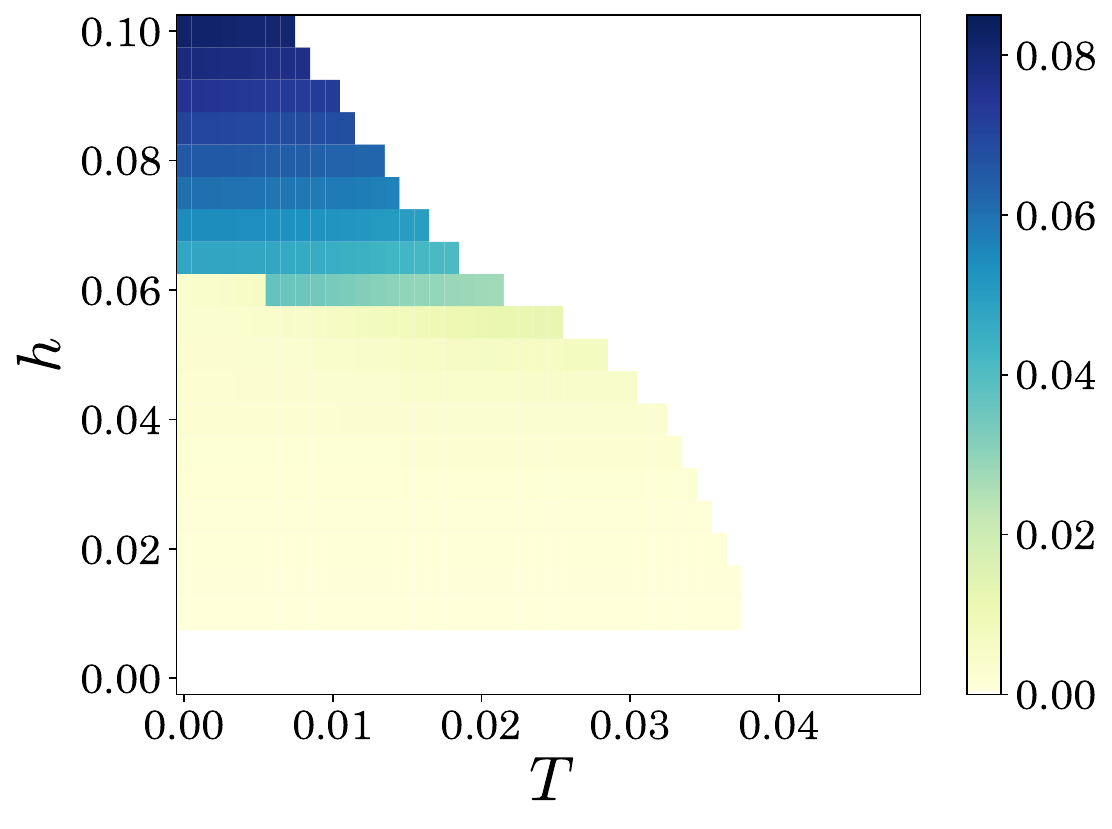}
\includegraphics[width=0.3\linewidth]{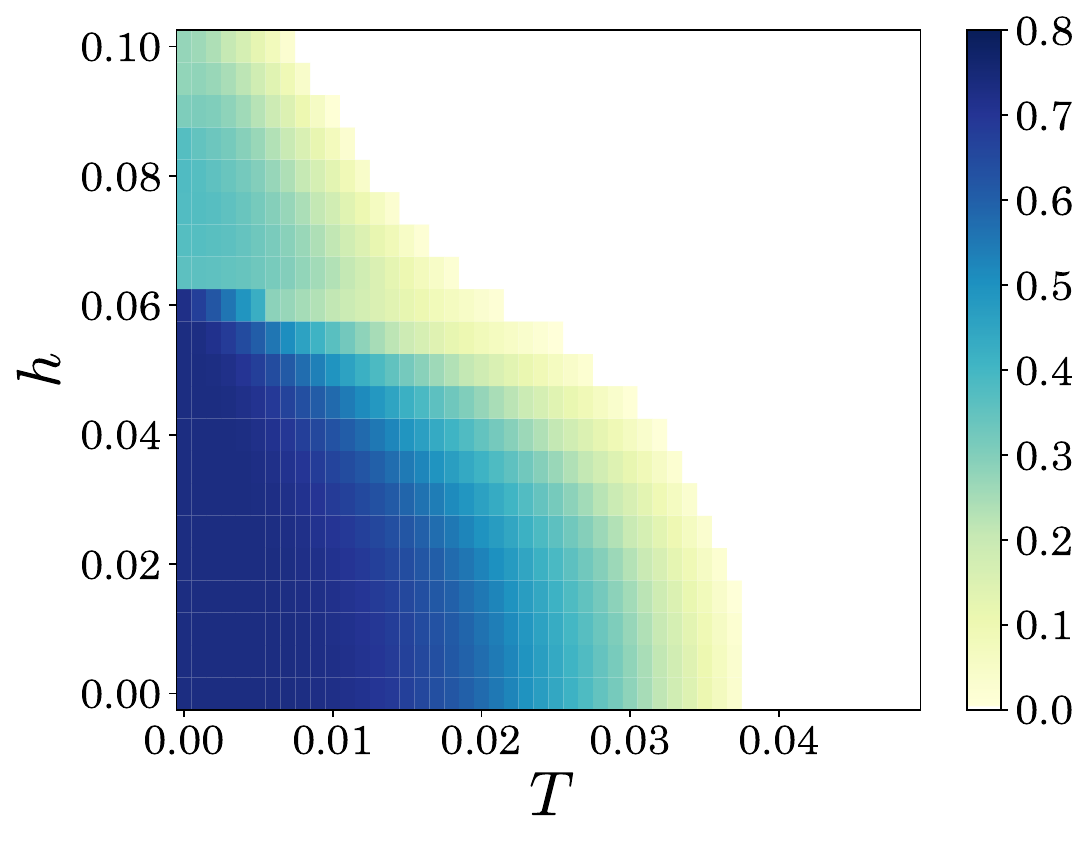}
\includegraphics[width=0.31\linewidth]{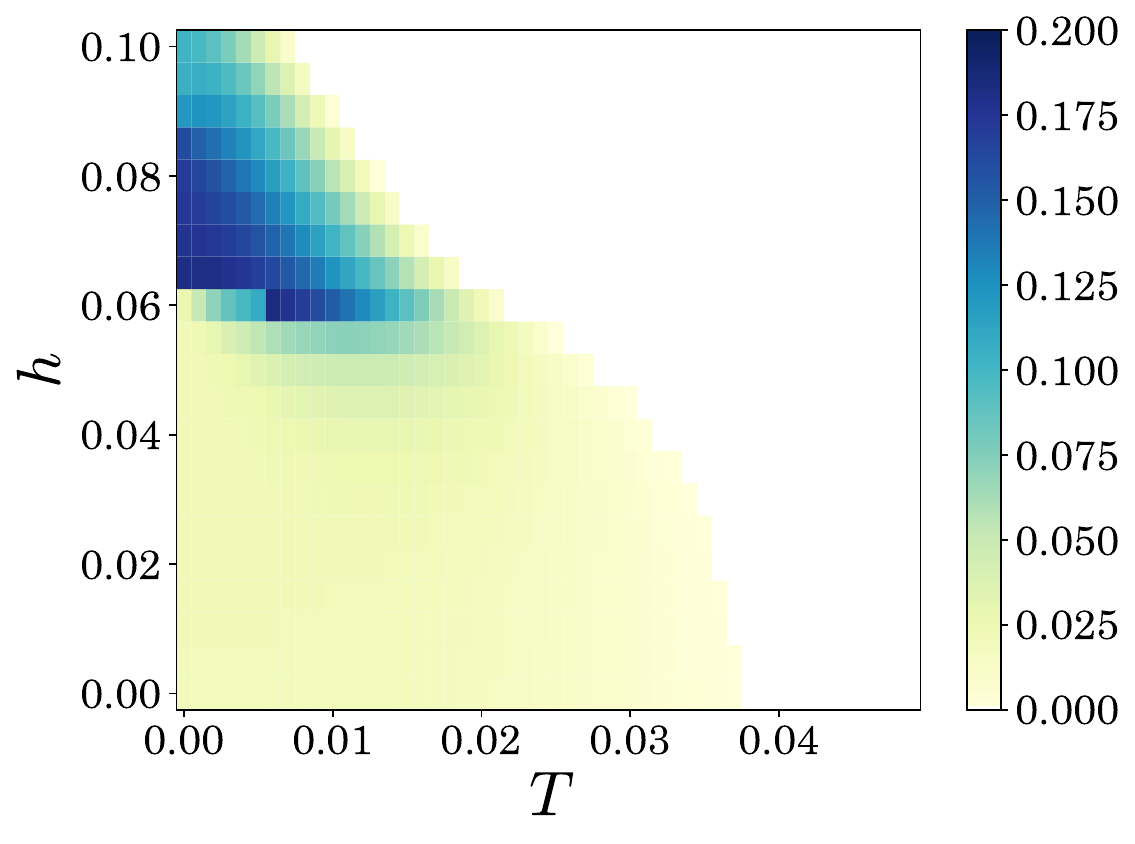}
\caption{The dependence of the COM momentum $\tilde{q}_0$ (Left), the superfluid weight $K_{xx}$ (Middle), and the spatially-dispersive superfluid weight $-K_{zxx}$ (Right) for an in-plane magnetic field $h$ and the temperature $T$. We use $t=1.0$, $\alpha = 0.3$, $\mu = -1.0$, and $U = 1.5$ for the numerical calculation. We use the numerical mesh $L_x = 10^6$ and $L_y = 200$, and the momentum is fixed to $2\pi n_x/L_x $ ($n_x \in \mathbb{Z}$) due to the periodic boundary condition.}  \label{phase_color_fig}
\end{figure*}
The normal Hamiltonian used in this letter is
\begin{subequations}
    \begin{gather}
        H = \sum_{\bm{k} s s'} H_{ss'\bm{k}} c^{\dagger}_{s\bm{k}} c_{s'\bm{k}} + H_{\mathrm{int}}, \\
        H_{ss' \bm{k} } = \xi_{\bm{k}} \delta_{ss'} + \bm{g}_{\bm{k}} \cdot \bm{\sigma}_{ss'} - \bm{h} \cdot \bm{\sigma}_{ss'}, \\
        H_{\mathrm{int}} = - \frac{U}{V} \sum_{\bm{k}\bm{k}'\tilde{\bm{q}}} c^{\dagger}_{\up\bm{k}^+} c^{\dagger}_{\down-\bm{k}^-} c_{\down-\bm{k}'^-} c_{\up\bm{k}'^+}.
    \end{gather}
\end{subequations}
Here, $H_{\bm{k}}$ is a two-dimensional one-band tight-binding model Hamiltonian including the spin-independent energy dispersion $\xi_{\bm{k}} = -2t(\cos k_x + \cos k_y) - \mu$ with the chemical potential $\mu$, the Rashba spin-orbit coupling (SOC) $\bm{g}_{\bm{k}} = \alpha ( -\sin k_y, \sin k_x ,0 )$, and the Zeeman coupling at a finite magnetic filed $\bm{h} \equiv - g_S \mu_B\bm{B} /2$ ($\mu_B$ is the Bohr magneton and $g_S \approx 2$ is the spin $g$-factor). $\bm{\sigma}$ are the Pauli matrices on the spin degrees of freedom. $c^{(\dagger)}_{s\bm{k}}$ is the annihilation (creation) operator with the spin $s=\up,\down$ and the wavenumber $\bm{k}$. $H_{\mathrm{int}}$ is the interaction Hamiltonian with the attractive interaction strength $U$, and $V$ is the volume of the system. We use the abbreviation $\bm{k}^{\pm} = \bm{k} \pm \tilde{\bm{q}}/2$.
Furthermore, we assume that the superconducting state is the spin-singlet pairing state with a finite COM momentum $\tilde{\bm{q}}$, and the order parameter is defined by
\begin{equation} \label{gap}
    \Delta_{\tilde{\bm{q}}} = - \frac{U}{V} \sum_{\bm{k}} \braket{ c_{\down -\bm{k}^-} c_{\up \bm{k}^+} }.
\end{equation}
The COM momentum can become finite when a supercurrent is finite in the equilibrium state or spontaneously becomes finite even in the absence of a supercurrent.
Particularly with regard to the latter, when applying an in-plane magnetic field (in our model, $\bm{h}$ is oriented in the $y$-direction), the COM of the Cooper pairs is finite and shifts along the perpendicular in-plane direction ($x$-direction in this model) due to the deformation of the Fermi surface, even without a supercurrent \cite{AGTERBERG200313,dimitrova2003phase,PhysRevLett.94.137002,PhysRevB.75.064511}. 
We use the mean field approximation and the Hamiltonian reads
\begin{subequations}
    \begin{gather}
        H = \frac{1}{2} \sum_{\bm{k}nm} \psi^{\dagger}_{n\bm{k},\tilde{\bm{q}}} H^{\mathrm{BdG}}_{nm\bm{k},\tilde{\bm{q}}} \psi_{m\bm{k}, \tilde{\bm{q}} } + \mathrm{const.} , \label{mean_field_hamiltonian} \\
        H^{\mathrm{BdG}}_{\bm{k},\tilde{\bm{q}}}
        =
        \begin{pmatrix}
            H_{\bm{k}^+} & \Delta_{\tilde{\bm{q}}} i \sigma_y \\
            - \Delta^{*}_{\tilde{\bm{q}}}i\sigma_y & - H^{\mathrm{T}}_{-\bm{k}^-}
        \end{pmatrix}.
    \end{gather}
\end{subequations}
Here, we define $O^\mathrm{T}$ as the transpose of a matrix $O$. $H^{\mathrm{BdG}}_{\bm{k},\tilde{\bm{q}}}$ is called the Bogoliubov-de Gennes (BdG) Hamiltonian, and the second term in Eq.~(\ref{mean_field_hamiltonian}) is a constant energy including the condensation energy by $\Delta_{\tilde{\bm{q}}}$. $\bm{\psi}^{\dagger}_{\bm{k},\tilde{\bm{q}}} = (c^{\dagger}_{\up \bm{k}^+}, c^{\dagger}_{\down \bm{k}^+}, c_{\up -\bm{k}^-}, c_{\down -\bm{k}^-})$ is the Nambu spinor.
Using this mean-field Hamiltonian, we solve the gap equation (\ref{gap}) at each $\tilde{\bm{q}}$ and denote the solution by $\bar{\Delta}_{\tilde{\bm{q}}}$. This solution provides the free energy density $F[\tilde{\bm{q}},h,T] = F(\bar{\Delta}_{\tilde{\bm{q}}},\tilde{\bm{q}},h,T)$ at each $\tilde{\bm{q}}$, fully describing the thermodynamic property of the equilibrium state (see supplemental materials for the detailed form $F(\cdots)$ \cite{suppl}). Especially, in the following, we denote the COM momentum under no supercurrent flow by $\tilde{\bm{q}}_0$, i.e., this momentum minimizes the free energy density: $\bm{j}(\bm{\tilde{\bm{q}}}_0, h , T ) = \partial F[\tilde{\bm{q}}_0,h,T]/ \partial \tilde{\bm{q}}_0  = 0$.

\textit{Superfluid weight.}---
The superfluid weight is encoded in the Meissner Kernel $K_{\alpha\beta}(\bm{q}) ( = \Phi_{\alpha\beta}(\bm{q}) - D_{\alpha \beta})$, where $\Phi_{\alpha \beta}(\bm{q})$ is the static current-current correlation function and $D_{\alpha \beta}$ is the Drude weight. 
Here, we consider the limitation to small $\bm{q}$ and expand the Meissner Kernel $K_{\alpha \beta}(\bm{q})$ by $\bm{q}$. The zeroth-order term $K_{\alpha \beta} = -K_{\alpha \beta}(0) =4e^2 \partial^2 F[\tilde{\bm{q}},h,T]/\partial \tilde{q}_{\alpha} \partial \tilde{q}_{\beta}$ is the usual superfluid weight, which is finite only in the superconducting state. The first-order term $K_{\alpha \beta \gamma} = -i \partial_{q_{\gamma}} K_{\alpha \beta}(0)$ represents the spatially dispersive effect of the superfluid weight. It also vanishes in the normal states \cite{PhysRevLett.116.077201}, and it is only finite in noncentrosymmetric superconductors. Thus, $K_{\alpha \beta \gamma}$ is one of the fundamental quantities characterizing noncentrosymmetric superconductors. 
It originates from the electromagnetic coupling, resulting in, e.g., the superconducting Edelstein effect \cite{levitov1985magnetoelectric,edel1989characteristics,PhysRevLett.75.2004,VictorMEdelstein_1996,PhysRevB.65.144508,PhysRevB.70.104521,PhysRevB.72.024515,PhysRevB.102.214510,he2019spin,PhysRevResearch.2.012073,PhysRevResearch.3.L032012,PhysRevLett.128.217703,PhysRevB.109.L220505} and even the modulation of the superconducting phase, such as the helical superconductor \cite{AGTERBERG200313,dimitrova2003phase,PhysRevLett.94.137002,PhysRevB.75.064511,PhysRevB.76.014522,BauerSigrist2012}. 

$K_{\alpha \beta}$ and $K_{\alpha \beta \gamma}$ are real, and they are symmetric and antisymmetric for the interchange of indices $\alpha$ and $\beta$, respectively \cite{PhysRevB.108.165119}. Especially, $K_{\alpha \beta \gamma}$ is directly related to the superconducting Edelstein coefficients $\mathcal{K}_{\alpha \beta} = e g_S \mu_{B} \partial^2 F[\tilde{\bm{q}},h,T]/\partial \tilde{q}_{\alpha} \partial h_{\beta}$ (here, $M_{\beta} = \mathcal{K}_{\alpha \beta} \delta \tilde{q}_{\alpha}/2$) as $K_{\alpha \beta \gamma} = \varepsilon_{\alpha \gamma \delta} \mathcal{K}_{\beta \delta} - \varepsilon_{\beta \gamma \delta} \mathcal{K}_{\alpha \delta}$, thus $K_{\alpha \beta \gamma}$ is experimentally detectable by e.g. the measurement of the supercurrent-induced magnetization and the missing area measurement of the optical activity proposed recently \cite{PhysRevB.108.165119,PhysRevB.110.085162}.

Here, we will evaluate $K_{\alpha \beta \gamma}$ to reveal how it characterizes the noncentrosymmetric superconductor. In this letter, we will study the helical superconductor with a spontaneously finite momentum of the Cooper pairs.
In our model given above, the magnetic point group of the system is $m'm2'$. Under this symmetry, the permitted components of the superfluid weights are $K_{xx}$, $K_{yy}$, $K_{zxx} = \mathcal{K}_{xy}$, and $K_{zyy} = -\mathcal{K}_{yx}$. 
In the following, we will focus on $K_{xx}$ and $K_{zxx}$. 
Although the superfluid weights can be derived from the second derivative of the free energy density by $\tilde{\bm{q}}$ or $\bm{h}$ discussed above, we will approximate them using a linear response formula without collective mode effects instead \cite{suppl}.

\textit{Phase diagram.}---
We show the dependence of the COM momentum $\tilde{q}_0$ on the magnetic filed $h$ and the temperature $T$ in Fig.~\ref{phase_color_fig} (left). We use the same parameter as used in Ref.~\cite{PhysRevLett.128.037001} and we reproduce the same phase diagram. The momentum $\tilde{q}_0$ is finite even for a small in-plane magnetic field as long as it is not zero. With increasing the field, the momentum jumps and is largely enhanced around $h = 0.06$. This jump is explained by the jump of the minimum of the free energy density \cite{PhysRevLett.128.037001}.
In Fig.~\ref{phase_color_fig} (middle) and (right), we show the superfluid weight $K_{xx}$ and the spatially-dispersive superfluid weight $K_{zxx}$, respectively. The value of each quantity rapidly changes around $h = 0.06$ along with the $\tilde{q}_0$'s jump. At this point, $K_{xx}$ is suppressed, whereas $K_{zxx}$ is enhanced. 

This sudden change of these quantities is attributed to the change of the Bogoliubov band (The band structure from the eigenenergy of the BdG Hamiltonian) near $E=0$ along with the $\tilde{q}_0$'s jump, as discussed in the following.
\begin{figure}[t]
\includegraphics[width=0.49\linewidth]{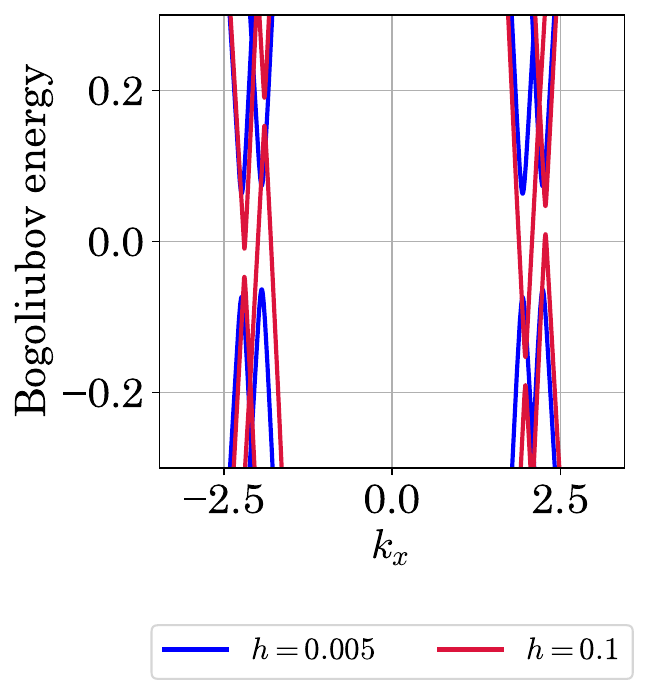}
\includegraphics[width=0.49\linewidth]{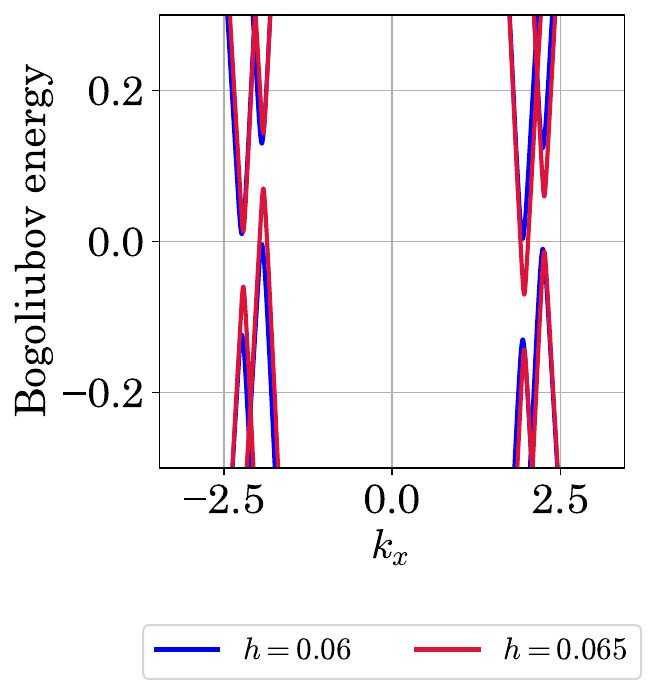}
\caption{The Bogoliubov band at $h=0.005$, $0.1$ (left), $0.06$, and $0.065$ (right). We set $k_y = 0$.}  \label{band}
\end{figure}
In Fig.~\ref{band}, we plot the Bogoliubov band at $k_y = 0$. The left figure shows the band structure for the low magnetic field $h=0.005$ and the high one $h=0.01$. For the small field, $h=0.005$, the band structure is similar to the zero-field case with a full gap at $E=0$. 
On the other hand, for the large magnetic field, the spectrum is gapless, which is also pointed out in Refs.~\cite{doi:10.1073/pnas.2019063118,PhysRevLett.128.037001}, because the gap points shift up or down due to the Doppler's shift by the COM momentum.
In this case, the band crosses $E=0$, leading to the formation of a structure resembling the Fermi surface, which is called the Bogoliubov Fermi surface (BFS).
The Doppler's shift scenario is well known in the B-phase of $^3\mathrm{He}$ with an applied supercurrent above the Landau critical current \cite{10.1093/acprof:oso/9780199564842.001.0001}.
It is also studied in electric systems \cite{doi:10.1073/pnas.2019063118,PhysRevB.97.115139,papaj2021creating,doi:10.1126/science.abf1077}.
Furthermore, we focus on the band structure immediately before and after the $\tilde{q}_0$'s jump in Fig.~\ref{band} (right). Just before the jump ($h=0.06$), the spectrum is still gapped, with the band edge approaching very close to $E=0$, as shown in Fig.~\ref{band} (right).
Just after the jump ($h=0.065$), the spectrum becomes gapless and forms the BFS. Thus, this fact indicates the $\tilde{q}_0$'s jump and the formation of the BFS occur simultaneously, as was also observed in the phase diagram in Ref.~\cite{PhysRevLett.128.037001}.
\begin{figure}[t]
\includegraphics[width=1\linewidth]{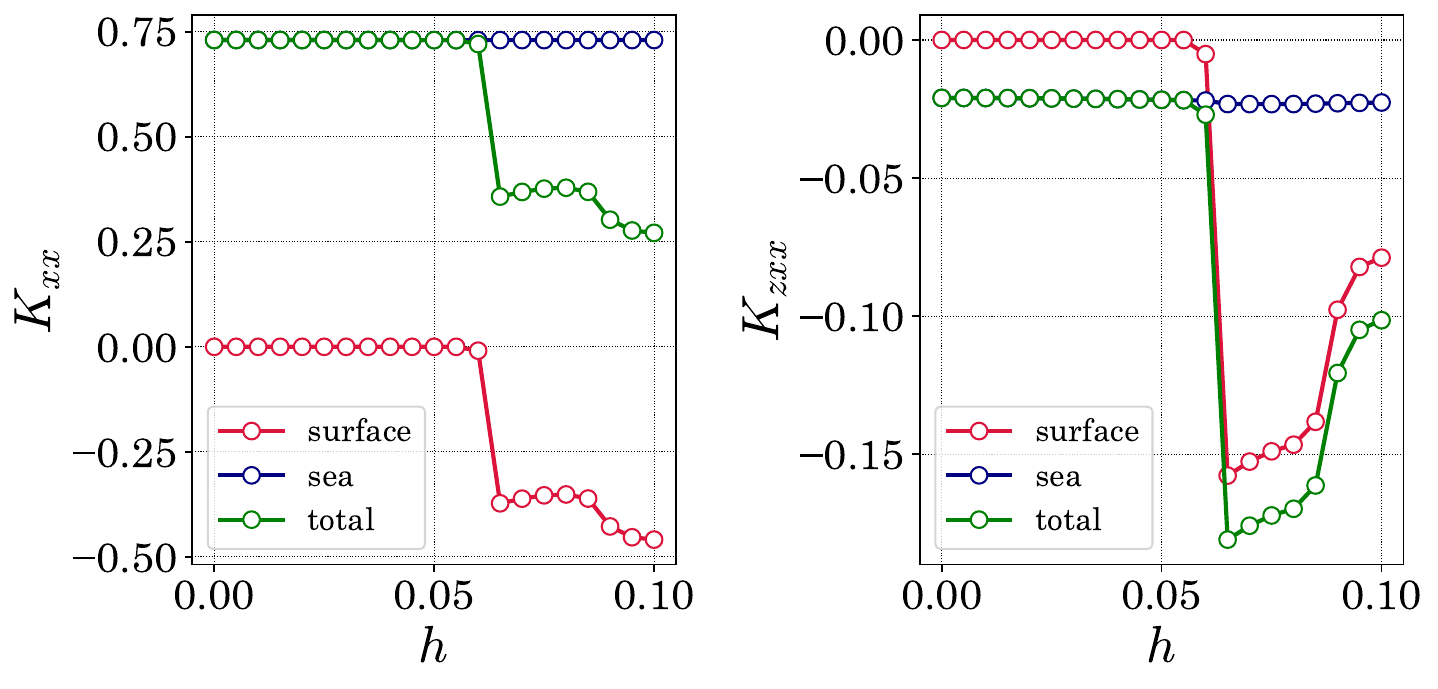}
\caption{The superfluid weights $K_{xx}$ (left) and $K_{zxx}$ (right) divided into the two effects, Bogoliubov Fermi surface term and the sea term, at $T=10^{-3}$.}  \label{sfd_surface_sea}
\end{figure}

This formation of the BFS largely affects the superfluid weights. In Fig.~\ref{sfd_surface_sea}, the superfluid weights are shown divided into two contributions, the Bogoliubov Fermi surface term and the sea term (see the supplemental materials for the detailed expressions of both terms \cite{suppl}). 
For all fields, the sea term is constant and is not strongly affected by the magnetic field.
At low fields, the sea term is dominant, and, in contrast, the surface term vanishes because the spectrum is gapped.
For large magnetic fields with BFS, the surface term largely contributes, reducing $K_{xx}$ and enhancing $K_{zxx}$.
The reason for the reduction of $K_{xx}$ is the finite quasiparticle density at $E=0$, resulting in a finite normal density and a reduction of the superfluid density. Although the superfluid density is also responsible for the finite $K_{zxx}$, its counterintuitive enhancement is surprising, indicating that it can capture other properties characterizing the superconducting state such as the COM momentum.

\begin{figure*}[t]
\includegraphics[width=0.32\linewidth]{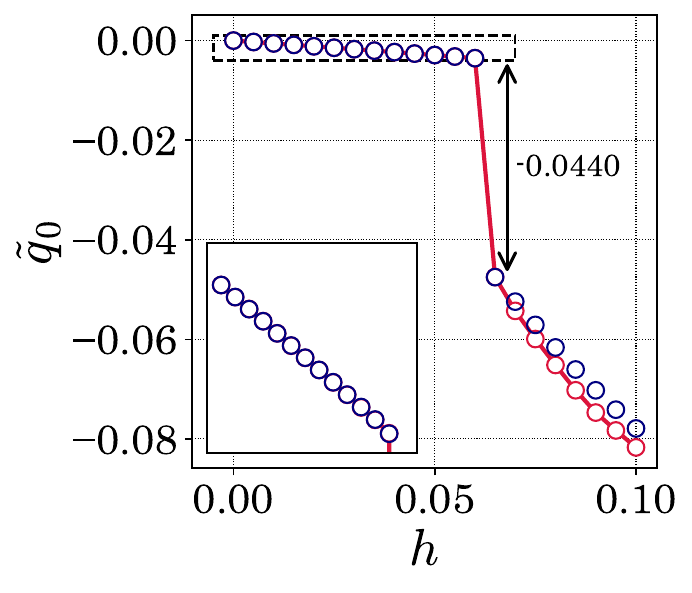}
\includegraphics[width=0.28\linewidth]{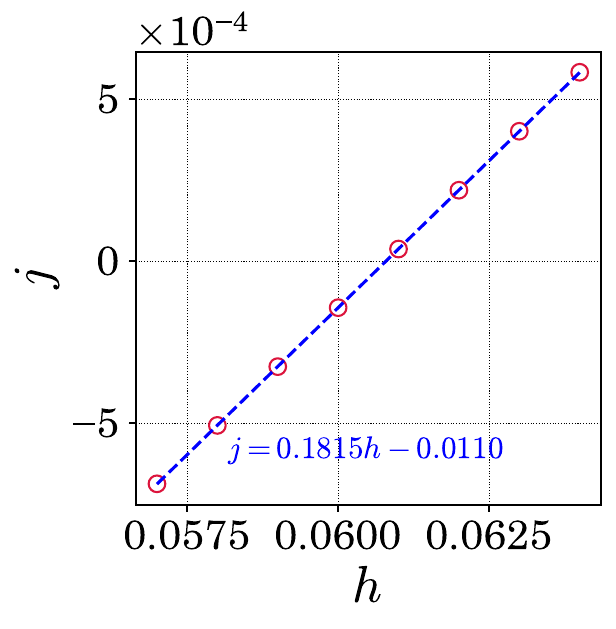}
\includegraphics[width=0.3\linewidth]{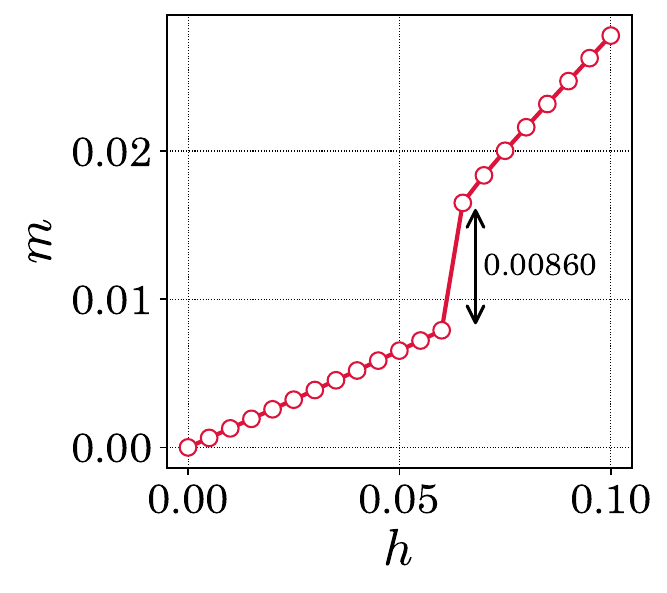}
\caption{(Left) The magnetic field dependence of the COM momentum $\tilde{q}_0$ at $T=10^{-3}$ (red). The result of integrating the thermodynamic relation is also plotted (blue). (Middle) The first-order transition line on the $j$-$h$ plane at $T=10^{-3}$. (Right) The magnetic field dependence of the magnetization along $y$-direction at $T=10^{-3}$.}  \label{jhphase_fig}
\end{figure*}

\textit{Thermodynamic relations.}---
Interestingly, the combination of the superfluid weights $K_{xx}$ and $K_{zxx}$ is known to give a direct relation with the COM momentum $\tilde{q}_0$ at low magnetic fields using the condition of the no supercurrent flow
\cite{AGTERBERG200313,PhysRevLett.94.137002,annurev:/content/journals/10.1146/annurev-conmatphys-031113-133912} as \footnote{In the following, we will denote by $K_{zxx}$ the original definition of $K_{zxx}$ divided by $-g_s \mu_B/2$.}
\begin{equation}
    \frac{\tilde{q}_0}{2} \overset{h \sim 0}{=} \frac{K_{zxx}(h=0) }{K_{xx}(h=0)} h. \label{Kaur_eq}
\end{equation}
We can generalize this argument. Under the constraint that no supercurrent flows, when the magnetic field varies, the change of the momentum $\tilde{q}_0$ is determined accordingly. Therefore, we obtain one exact thermodynamic relation valid in the entire regime of the phase diagram as \cite{suppl}
\begin{equation}
    \frac{\delta \tilde{q}_0}{\delta h} = \frac{2 K_{zxx}}{K_{xx}}. \label{thermo_relation1}
\end{equation}
This relation is the first main result of this letter.
The quantity on the right-hand side can be experimentally measured, thus the COM momentum in the phase diagram can be precisely obtained. We note that the momentum $\tilde{q}_0$ is zero at zero magnetic field due to the time-reversal symmetry.
Here, we focus on our model case (the two-dimensional Rashba system), however, similar relations hold for all other helical superconductors \cite{suppl}.

This relation has one problem: it cannot track the momentum in cases where discontinuous changes, such as first-order phase transitions, occur.
As seen above, such a situation also occurs in the current model. However, this discontinuous jump of the momentum can be measured by using another thermodynamic relation analogous to the Clausius–Clapeyron equation known in the liquid-gas transition.
We introduce the conjugated free energy $G[j,h,T] = \min_{\tilde{q}}( F[\tilde{q},h,T] - j \tilde{q} )$ by the Legendre transformation. $j$ is the supercurrent conjugated to the momentum $\tilde{q}$ and serves as a control parameter in the experiment.
Now, let us consider points $(h_{*},j_{*})$ along the first-order transition line in the $j$-$h$ plane. From the continuity of the $G$ function, we obtain the second thermodynamic relation \cite{suppl}
\begin{equation}
    \frac{d j_{*}}{d h_{*}} = -\frac{\Delta m}{\Delta \tilde{q}}. \label{CC_relation}
\end{equation}
This equation means that the slope of the transition line is determined by the ratio of the jumps of the magnetization $\Delta m$ and the momentum $\Delta \tilde{q}$. This slope and the magnetization are observable, thus, we can experimentally determine the jump of the momentum using this relation, providing strong evidence of the spontaneously finite COM momentum using data in the vicinity of $j_* = 0$.
Instead of considering the magnetic field, we can also consider the temperature $T$.
In this case, $\Delta m$ is replaced by the jump of the entropy $\Delta s$.

In Fig.~\ref{jhphase_fig} (left), we plot the magnetic field dependence of $\tilde{q}_0$ at $T=10^{-3}$ (red). The momentum $\tilde{q}_0$ jumps around $h=0.06$, and the jump value $\Delta \tilde{q}_0$ is $-0.0440$. In the regions before and after the transition, the momentum can be obtained by integrating the ratio of the superfluid weights along the magnetic field in Eq.~(\ref{thermo_relation1}) respectively. Assuming the initial values at $h=0$ and $h=0.065$ are consistent with the actual momentum $\tilde{q}_0$, the integrated results (blue) are in good agreement with $\tilde{q}_0$.
There is a slight deviation for high fields. This is a practical problem of the calculation \cite{suppl}, and we note that the two values should be consistent due to the thermodynamic relation in Eq.~(\ref{thermo_relation1}).

In Fig.~\ref{jhphase_fig} (middle), the first-order transition line on the $j$-$h$ plane in the vicinity of $j=0$ is plotted, and the slope is 0.182. 
The line corresponds to the non-differentiable points of $G[j,h,T]$ \cite{suppl}.
In Fig.~\ref{jhphase_fig} (right), the magnetic field dependence of the magnetization is plotted, and it shows the jump around $h=0.06$ with $\Delta m = 0.00860$. Using the thermodynamic relation in Eq.~(\ref{CC_relation}), the momentum jump $\Delta \tilde{q}_0$ is converted as $-\Delta m / (dj_* / dh_*) = -0.0473$, which is almost consistent with the true value.
We note that there is no guarantee that the thermodynamic relation holds when using free energy with the mean-field approximation, which may destroy the thermodynamic properties such as a convexity. However, this is a practical obstacle, and the relation should be established.

\begin{figure}[t]
\includegraphics[width=0.8\linewidth]{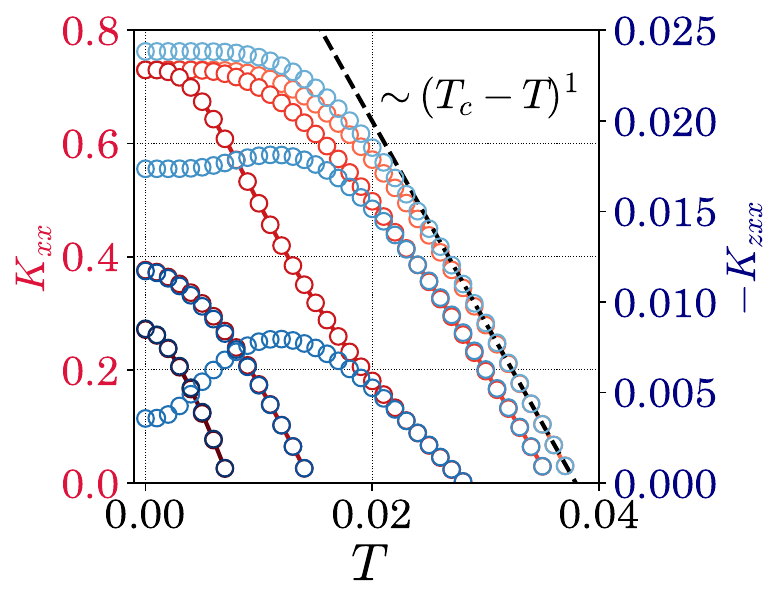}
\caption{The temperature dependence of $K_{xx}$ (red) and $K_{zxx}$ (blue) at some magnetic fields. The curves for $h=0.005$, $0.025$, $0.05$, $0.075$, and $0.1$ are in order from the highest transition temperature. $K_{zxx}$ is shown multiplied by each constant factor: 1.135, 0.82, 0.165, 0.068, and 0.084 for each curve.}  \label{temp_dep_fig}
\end{figure}

\textit{Temperature dependence.}---
\begin{figure}[b]
\includegraphics[width=0.48\linewidth]{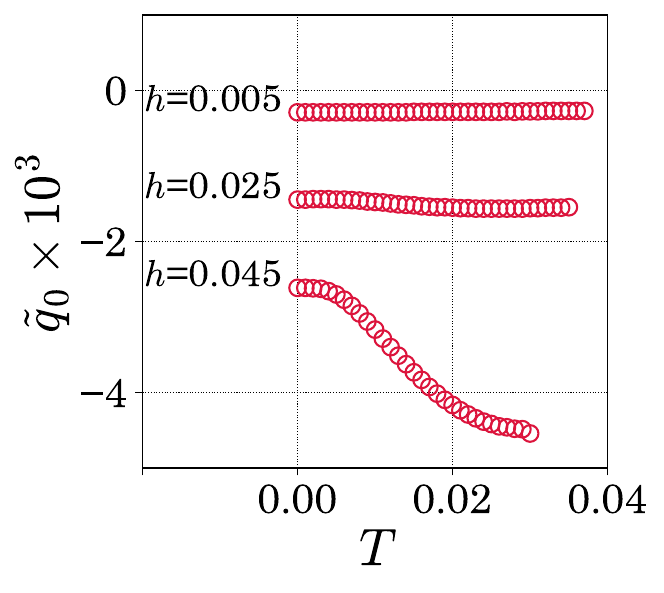}
\includegraphics[width=0.5\linewidth]{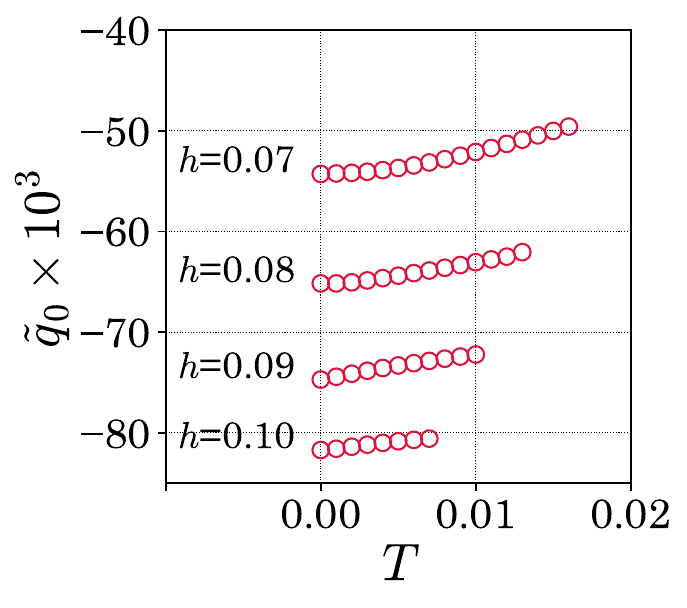}
\caption{The temperature dependence of the COM momentum $\tilde{q}_0$ at low magnetic fields $h = 0.01, 0.03$, and $0.05$ (Left) and at high fields $h=0.07$, $0.08$, $0.09$, and $0.1$ (Right).}  \label{comp_momentum_highmag_fig}
\end{figure}
In Fig.~\ref{comp_momentum_highmag_fig}, we show the temperature dependence of the COM momentum $\tilde{q}_0$ at low magnetic fields (left) and high fields (right).
One feature is that the momentum remains finite in the vicinity of the transition temperature $T_c$. 
This feature is explained by the temperature dependence of the ratio of the two superfluid weights $K_{xx}$ and $K_{zxx}$.
They fall continuously to zero around $T_c$ as shown in Fig~\ref{temp_dep_fig}.
However, these quantities display the same temperature dependence around $T_c$, where they decrease linearly to zero ($\sim (T_c - T)^1$). Thus, the ratio approaches a certain constant due to the cancellation of the temperature dependence around $T_c$ and is finite and has a constant sign at all points on the phase diagram. It reproduces the COM momentum dependence using the thermodynamic relation (Eq.~\ref{thermo_relation1}).

This linear behavior of $K_{xx}$ and $K_{zxx}$ can be explained using Ginzburg-Landau (GL) theory at low magnetic fields. The GL free energy density for the order parameter $\Delta$ (assumed real) with the dispersion along the $x$-direction $\tilde{q}$ under magnetic field reads
\begin{equation}
    f(\Delta, \tilde{q}, h) = - a(\tilde{q}) \Delta^2 + \frac{b}{2} \Delta^4, 
\end{equation}
where $a = a_0 + a_1 \tilde{q} h + a_2 \tilde{q}^2$. $a_0$ is assumed to be proportional to $(T_c -T)^1$, and $a_1$ is unique for systems without an inversion center. It corresponds to the electromagnetic coupling in our model, where we apply the in-plane magnetic field $h$ along the $y$-direction. 
In addition, we assume that the magnetic effect only induces the Zeeman effect because of the suppression of the orbital effect in two-dimensional systems. 
The order parameter minimizing the GL free energy density $\bar{\Delta}$ satisfies $\partial f/\partial \bar{\Delta} = 0 $, and we get $\bar{\Delta}^2 = a(\tilde{q})/b$. Then, the free energy density of this solution is $f[\tilde{q},h] =  f(\bar{\Delta}, \tilde{q}, h) = - a(\tilde{q})^2/2b$. 
Furthermore, the COM momentum is determined by the no current condition $\partial f[\tilde{q}_0,h]/\partial \tilde{q}_0 = 0$, and we obtain the solution $\tilde{q}_0 = -a_1 h / 2a_2$, demonstrating that the helical phase originates in the electromagnetic coupling $a_1$ under a magnetic field \cite{PhysRevLett.94.137002}.
The ratio of $a_1$ to $a_2$ can be expressed as the ratio of the superfluid weights at zero magnetic field.
Actually, the superfluid weights are given by
\begin{subequations}
    \begin{align}
        K_{xx}(h=0) &\propto \frac{\partial^2 f}{\partial \tilde{q}^2}\biggr|_{\substack{h \to 0 \\ \tilde{q} \to 0 }} = - \frac{a_0 a_2}{b} \propto (T_c -T)^1, \\
        K_{zxx}(h=0) &\propto \frac{\partial^2 f}{\partial h \partial \tilde{q}}\biggr|_{\substack{h \to 0 \\ \tilde{q} \to 0 }} = - \frac{a_0 a_1}{b} \propto (T_c -T)^1.
    \end{align}
\end{subequations}
Thus, $a_1/a_2$ is equivalent to $K_{zxx}(h=0)/K_{xx}(h=0)$ except for an universal constant.
These equations show that the superfluid weights have a linear dependence on the temperature, resulting in the temperature independence of $\tilde{q}_0$ for low magnetic fields.

\textit{Summary and Discussion.}---
We have proposed two general thermodynamic relations applicable to all helical superconductors for detecting the COM momentum. 
The first relation determines the derivative of the COM momentum by the magnetic field as the ratio of the superfluid weight and the superconducting Edelstein coefficient. The ratio is experimentally measured by the superconducting Edelstein effect and it provides the COM momentum in the entire region of the phase diagram. 
The other is related to the first-order transition, where the jump of the momentum is determined by the slope of the transition line generalized from the Clausius–Clapeyron equation. This relation alone provides strong evidence for the finite COM momentum, and its verification seems relatively easy because it only requires obtaining the phase diagram under a supercurrent.

Finally, we evaluate the order of the magnitude of the induced spin moment for future experiments. The supercurrent-induced spin moment is $\delta s_y / \mu_{B} = K_{zxx}/K_{xx} J^{\mathrm{sc}}_x$. The coefficient is $K_{zxx}/K_{xx} = 0.029 \hbar/eta$ at $T=10^{-3}$ for zero magnetic field. For instance, when we set $t=1~\mathrm{eV}$ and $a = 1~\mathrm{\AA}$, we obtain $K_{zxx}/K_{xx} = 0.12 \times 10^{13}~\mathrm{A}^{-1} \mathrm{m}^{-1}$. 
This ratio is hardly scaled by the superconducting gap and is not so different from the value in the normal phase.
In addition, when we apply a supercurrent $J^{\mathrm{sc}}_x = 10^{-3}\mathrm{-}10^{0}~\mathrm{A}\mathrm{m}^{-1}$, the induced spin moment is $0.12\times 10^{-8}\mathrm{-}10^{-5}~\mathrm{nm}^{-2}$. Above the magnetic field where the COM momentum $\tilde{q}$ jumps, the induced moment is ten times larger. This spin moment can be observed by e.g. the spin-polarised positron beam \cite{PhysRevLett.49.1779,KAWASUSO2013139,PhysRevLett.114.166602} and the state-of-the-art sensor sensitive to weak magnetic fields such as SQUID and NV-center. We note that another direct measurement of the spin moment is recently proposed, including the measurement of an additional phase in a Josephson junction \cite{senapati2023phase}. Furthermore, we can use the missing area measurement of the optical activity to measure the superconducting Edelstein coefficient \cite{PhysRevB.108.165119,PhysRevB.110.085162}.

K.S. appreciates helpful discussions with H. Tanaka.
K.S. acknowledges support as a JSPS research fellow and is supported by JSPS KAKENHI, Grant No.22J23393 and No.22KJ2008. 
R.P. is supported by JSPS KAKENHI No.23K03300.

\bibliography{reference.bib}

\clearpage
\renewcommand{\bibnumfmt}[1]{[S#1]}
\renewcommand{\citenumfont}[1]{S#1}
\renewcommand{\thesection}{S\arabic{section}}
\renewcommand{\theequation}{S\arabic{equation}}
\setcounter{equation}{0}
\renewcommand{\thefigure}{S\arabic{figure}}
\setcounter{figure}{0}
\renewcommand{\thetable}{S\arabic{table}}
\setcounter{table}{0}
\makeatletter
\c@secnumdepth = 2
\makeatother
\onecolumngrid
\begin{center}
 {\large \textmd{Supplemental Materials:} \\[0.3em]
 {\bfseries Thermodynamic relations for the Cooper pair's momentum in helical superconductors}}
\end{center}
\onecolumngrid

\section{linear response formula of superfluid weights and relation with the free energy density}
We introduce the band representation in the clean limit, and the static current-current correlation function reads
\begin{subequations}
\begin{align}
    \Phi_{\alpha \beta}(\bm{q}) &= 
    \frac{-1}{2} \sum_{nm} \int_{\bm{k}} \frac{f(E_{n\bm{k}_-,\tilde{\bm{q}}} ) -  f(E_{m\bm{k}_+,\tilde{\bm{q}}} ) }{E_{n\bm{k}_-,\tilde{\bm{q}}} - E_{m\bm{k}_+,\tilde{\bm{q}}}} W^{\alpha \beta}_{nm\bm{k},\tilde{\bm{q}}}(\bm{q}), \\
    W^{\alpha \beta}_{nm\bm{k},\tilde{\bm{q}}}(\bm{q}) &=
    \bra{ u_{n \bm{k}_- , \tilde{\bm{q}} } }  \tilde{j}^{\alpha}_{\bm{k},\tilde{\bm{q}},\bm{q}} \ket{ u_{m \bm{k}_+ , \tilde{\bm{q}} } }
    \bra{ u_{m \bm{k}_+ , \tilde{\bm{q}} } }  \tilde{j}^{\beta}_{\bm{k},\tilde{\bm{q}},-\bm{q}} \ket{ u_{n \bm{k}_- , \tilde{\bm{q}} } }.
\end{align}    
\end{subequations}
Here, we use the eigenstate $\ket{ u_{m \bm{k} , \tilde{\bm{q}}} }$ of the BdG Hamiltonian $H^{\mathrm{BdG}}_{\bm{k},\tilde{\bm{q}}}$ with the eigenenergy $E_{m\bm{k},\tilde{\bm{q}}}$, and the Fermi distribution function $f(x) = 1/(e^{\beta x} + 1)$ at the temperature $\beta^{-1}$. We define $\bm{k}_{\pm} = \bm{k} \pm \bm{q}/2$. We note that we have two kinds of momenta. One is the momentum of the Cooper pairs $\tilde{\bm{q}}$ and the other is the momentum representing the spatial dispersion of the superfluid density $\bm{q}$. We use the abbreviation $\pm$ for $\tilde{\bm{q}}$ by the superscript and $\bm{q}$ by the subscript. $\tilde{\bm{j}}_{\bm{k},\tilde{\bm{q}},\bm{q}}$ is the current density operator extended to the particle-hole space and is defined as
\begin{equation}
    \tilde{\bm{j}}_{\bm{k},\tilde{\bm{q}},\bm{q}} = 
    \begin{pmatrix}
        \bm{j}_{\bm{k}^+,\bm{q}} & 0 \\
        0 & -\bm{j}^{\mathrm{T}}_{-\bm{k}^-,\bm{q}}
    \end{pmatrix},
\end{equation}
using the normal current density operator $\bm{j}_{\bm{k},\bm{q}} = - e \bm{v}_{\bm{k}} - \frac{g_S \mu_B}{2} i \bm{q} \times \bm{\sigma}$, where $\bm{v}_{\bm{k}} = \partial H_{\bm{k}}/ \partial \bm{k}$.
The diamagnetic term extended to the particle-hole space is
\begin{subequations}
    \begin{gather}
        D_{\alpha \beta} = \frac{-e^2}{2} \sum_{n} \int_{\bm{k}} f(E_{n\bm{k},\tilde{\bm{q}}}) \bra{u_{n\bm{k},\tilde{\bm{q}}}}
        \tilde{v}^{\alpha \beta}_{\bm{k},\tilde{\bm{q}}}
        \ket{u_{n\bm{k},\tilde{\bm{q}}}}, \\
        \tilde{v}^{\alpha \beta}_{\bm{k},\tilde{\bm{q}}}
        =
        \begin{pmatrix}
        v^{\alpha \beta}_{\bm{k}^+} & 0 \\
        0 & -v^{\alpha \beta \mathrm{T}}_{-\bm{k}^-} 
        \end{pmatrix}
        =
        \frac{\partial^2 H^{\mathrm{BdG}}_{\bm{k},\tilde{\bm{q}}} }{\partial (\tilde{q}_{\alpha}/2) \partial (\tilde{q}_{\beta}/2)}.
    \end{gather}
\end{subequations}
Here, $v^{\alpha \beta}_{\bm{k}} = \partial^2 H_{\bm{k}}/\partial k_{\alpha} \partial k_{\beta}$.

Taking the limit of $\bm{q} \to 0$, we obtain the superfluid weight. We can derive the direct relation with the free energy density as seen in the following. The static correlation function is divided into the intraband term and interband term,
\begin{subequations}
    \begin{gather}
        \Phi^{(\mathrm{intra})}_{\alpha \beta}(0) 
        = -\frac{e^2}{2} \sum_n \int_{\bm{k}} \frac{\partial f(E_{n\bm{k},\tilde{\bm{q}}}) }{\partial E_{n\bm{k},\tilde{\bm{q}}}} 
        \mathrm{Re}[
        \bra{ u_{n \bm{k} , \tilde{\bm{q}} } }  \tilde{v}^{\alpha}_{\bm{k},\tilde{\bm{q}}} \ket{ u_{n \bm{k} , \tilde{\bm{q}} } }
        \bra{ u_{n \bm{k} , \tilde{\bm{q}} } }  \tilde{v}^{\beta}_{\bm{k},\tilde{\bm{q}}} \ket{ u_{n \bm{k} , \tilde{\bm{q}} } }
        ], \\
        \Phi^{(\mathrm{inter})}_{\alpha \beta}(0) 
        = -\frac{e^2}{2} \sum_{n\neq m} \int_{\bm{k}} \frac{f(E_{n\bm{k},\tilde{\bm{q}}} ) -  f(E_{m\bm{k},\tilde{\bm{q}}} ) }{E_{n\bm{k},\tilde{\bm{q}}} - E_{m\bm{k},\tilde{\bm{q}}}}
        \mathrm{Re}[
        \bra{ u_{n \bm{k} , \tilde{\bm{q}} } }  \tilde{v}^{\alpha}_{\bm{k},\tilde{\bm{q}}} \ket{ u_{m \bm{k} , \tilde{\bm{q}} } }
        \bra{ u_{m \bm{k} , \tilde{\bm{q}} } }  \tilde{v}^{\beta}_{\bm{k},\tilde{\bm{q}}} \ket{ u_{n \bm{k} , \tilde{\bm{q}} } }
        ].
    \end{gather}
\end{subequations}
Here, we define the velocity operator extended to the particle-hole space by
\begin{equation}
    \tilde{v}^{\alpha}_{\bm{k},\tilde{\bm{q}}}
    =
    \begin{pmatrix}
        v^{\alpha}_{\bm{k}^+} & 0\\
        0 & -v^{\alpha \mathrm{T}}_{-\bm{k}^-}
    \end{pmatrix}
    =
    \frac{\partial H^{\mathrm{BdG}}_{\bm{k},\tilde{\bm{q}}}}{\partial (\tilde{q}_{\alpha}/2)}.
\end{equation}
The intraband term originates from only the Bogoliubov Fermi surface. Thus, for usual superconductors, this term becomes zero at low temperatures because of the superconducting gap.
Using the relation $\bra{ u_{n \bm{k} , \tilde{\bm{q}} } }  \tilde{v}^{\alpha}_{\bm{k},\tilde{\bm{q}}} \ket{ u_{m \bm{k} , \tilde{\bm{q}} } } = ( E_{n\bm{k},\tilde{\bm{q}}} - E_{m\bm{k},\tilde{\bm{q}}} ) \braket{ \partial_{\tilde{q}_{\alpha}/2} u_{n \bm{k} , \tilde{\bm{q}} } | u_{m \bm{k} , \tilde{\bm{q}} }  } $ for $n \neq m$, we can rewrite the interband term as
\begin{align}
    \Phi^{(\mathrm{inter})}_{\alpha \beta}(0) 
        &=
        -\frac{e^2}{2} \sum_{n\neq m} \int_{\bm{k}} (f(E_{n\bm{k},\tilde{\bm{q}}} ) -  f(E_{m\bm{k},\tilde{\bm{q}}} ))
        \mathrm{Re}[
        \braket{ \partial_{\tilde{q}_{\alpha}/2} u_{n \bm{k} , \tilde{\bm{q}} } | u_{m \bm{k} , \tilde{\bm{q}} }  }
        \bra{ u_{m \bm{k} , \tilde{\bm{q}} } }  \tilde{v}^{\beta}_{\bm{k},\tilde{\bm{q}}} \ket{ u_{n \bm{k} , \tilde{\bm{q}} } }
        ] \nonumber \\
        &=
        -\frac{e^2}{2} \sum_{n\neq m} \int_{\bm{k}} 2 f(E_{n\bm{k},\tilde{\bm{q}}} ) 
        \mathrm{Re}[
        \braket{ \partial_{\tilde{q}_{\alpha}/2} u_{n \bm{k} , \tilde{\bm{q}} } | u_{m \bm{k} , \tilde{\bm{q}} }  }
        \bra{ u_{m \bm{k} , \tilde{\bm{q}} } }  \tilde{v}^{\beta}_{\bm{k},\tilde{\bm{q}}} \ket{ u_{n \bm{k} , \tilde{\bm{q}} } }
        ] \nonumber \\
        &=
        -\frac{e^2}{2} \sum_{n} \int_{\bm{k}} f(E_{n\bm{k},\tilde{\bm{q}}} ) 
        \mathrm{Re}[
        \bra{ \partial_{\tilde{q}_{\alpha}/2} u_{n \bm{k} , \tilde{\bm{q}} } }  \tilde{v}^{\beta}_{\bm{k},\tilde{\bm{q}}} \ket{ u_{n \bm{k} , \tilde{\bm{q}} } }
        +
        \bra{ u_{n \bm{k} , \tilde{\bm{q}} } }  \tilde{v}^{\beta}_{\bm{k},\tilde{\bm{q}}} \ket{ \partial_{\tilde{q}_{\alpha}/2}  u_{n \bm{k} , \tilde{\bm{q}} } }
        ] \nonumber \\
        &=
        -\frac{e^2}{2} \sum_{n} \int_{\bm{k}} f(E_{n\bm{k},\tilde{\bm{q}}} ) 
        \mathrm{Re}[
        \partial^2 E_{n\bm{k},\tilde{\bm{q}}}/ \partial (\tilde{q}_{\alpha}/2) \partial (\tilde{q}_{\beta}/2)
        -
        \bra{ u_{n \bm{k} , \tilde{\bm{q}} } }  \tilde{v}^{\alpha \beta}_{\bm{k},\tilde{\bm{q}}} \ket{ u_{n \bm{k} , \tilde{\bm{q}} } }
        ].
\end{align}
The last term cancels out the diamagnetic term $D_{\alpha \beta}$. Therefore, the total superfluid weight $K_{\alpha \beta} = -\Phi_{\alpha \beta}(0) + D_{\alpha \beta}$ is
\begin{align}
    K_{\alpha \beta}
    &=
    \frac{e^2}{2} \frac{\partial}{\partial (\tilde{q}_{\alpha} /2)  } \sum_n \int_{\bm{k}}
    f(E_{n\bm{k},\tilde{\bm{q}}}) \frac{\partial E_{\bm{k},\tilde{\bm{q}}}}{\partial (\tilde{q}_{\beta}/2)} \nonumber \\
    &=
    e^2 \frac{\partial^2 F(\Delta,\tilde{\bm{q}})}{\partial (\tilde{q}_{\alpha} /2) \partial (\tilde{q}_{\beta} /2) }.
\end{align}
Here, we define the free energy density by
\begin{align} \label{free_energy}
F(\Delta,\tilde{\bm{q}})
&=-\beta^{-1} \ln \mathrm{Tr} e^{- \beta H} \nonumber \\
&= - \frac{1}{2\beta}\sum_n \int_{\bm{k}} \ln(1 + e^{-\beta E_{n\bm{k},\tilde{\bm{q}}}}) + \frac{1}{2} \int_{\bm{k}} \mathrm{Tr} H_{\bm{k}} + \frac{|\Delta|^2}{U}.
\end{align}
Now, we note that the actual superfluid weight is expressed by the second derivative of $F[\tilde{\bm{q}}] = F(\Delta_{\tilde{\bm{q}}},\tilde{\bm{q}})$ with respect to $\tilde{\bm{q}}$, as $K_{\alpha \beta} = 4e^2 \partial^2 F[\tilde{\bm{q}}]/\partial \tilde{q}_{\alpha} \partial \tilde{q}_{\beta}$. Thus, the above linear response formula is an approximation of the superfluid weight neglecting the $\tilde{q}$ dependence of the gap parameter, which is expected to be restored by incorporating interaction effects through vertex corrections. 
Here, this correction is neglected, and the slight difference from the momentum $\tilde{q}_0$ in Fig.~\ref{jhphase_fig} may be caused by this approximation.

Next, we will derive the spatially dispersive superfluid weight $K_{\alpha \beta \gamma}$. Then, we need the derivative of Meissner Kernel $K_{\alpha \beta}(\bm{q})$ by $\bm{q}$. Two possible contributions can be considered. One is the spin part originating from the Zeeman term, and the other is the orbital part. In our model, the orbital part does not contribute because of the two-dimensional system, thus we will discuss only the spin part in the following. The spin part reads
\begin{align}
    K_{\alpha \beta \gamma} 
    &=
    -i \partial_{q_{\gamma}} K_{\alpha \beta}(0) \nonumber \\
    &=
    -\frac{e g_S \mu_B}{4} \sum_{nm} \int_{\bm{k}}
    \frac{f(E_{n\bm{k},\tilde{\bm{q}}} ) -  f(E_{m\bm{k},\tilde{\bm{q}}} ) }{E_{n\bm{k},\tilde{\bm{q}}} - E_{m\bm{k},\tilde{\bm{q}}}}
    \mathrm{Re}[
    \bra{ u_{n\bm{k},\tilde{\bm{q}}} } \varepsilon_{\alpha \gamma \delta} \tilde{\sigma}_{\delta} \ket{ u_{m\bm{k},\tilde{\bm{q}}} }
    \bra{ u_{m\bm{k},\tilde{\bm{q}}} } \tilde{v}^{\beta}_{\bm{k},\tilde{\bm{q}}} \ket{ u_{n\bm{k},\tilde{\bm{q}}} } \nonumber \\
    &-
    \bra{ u_{n\bm{k},\tilde{\bm{q}}} } \tilde{v}^{\alpha}_{\bm{k},\tilde{\bm{q}}} \ket{ u_{m\bm{k},\tilde{\bm{q}}} }
    \bra{ u_{m\bm{k},\tilde{\bm{q}}} } \varepsilon_{\beta \gamma \delta} \tilde{\sigma}^{\delta} \ket{ u_{n\bm{k},\tilde{\bm{q}}} }
    ].
\end{align}
Here, the spin operator extended to the particle-hole space is defined by
\begin{equation}
    \tilde{\sigma}^{\delta} 
    =
    \begin{pmatrix}
        \sigma^{\delta} & 0 \\
        0 & - \sigma^{\delta \mathrm{T}}
    \end{pmatrix}
    =
    -\frac{\partial H^{\mathrm{BdG}}_{\bm{k},\tilde{\bm{q}}} }{\partial h_{\delta}}.
\end{equation}
Dividing $K_{\alpha \beta \gamma}$ into the intraband and interband parts as
\begin{subequations}
\begin{align}
        K^{\mathrm{(intra)}}_{\alpha \beta \gamma}
        &=
        -\frac{eg_S \mu_B}{4} \sum_n \int_{\bm{k}} \biggl\{
        -\varepsilon_{\alpha \gamma \delta} \frac{\partial f(E_{n\bm{k},\tilde{\bm{q}}}) }{\partial (\tilde{q}_\beta/2)} \frac{\partial E_{n\bm{k},\tilde{\bm{q}}}}{\partial h_{\delta}} 
        -
        ( \alpha \leftrightarrow \beta )
        \biggr\},  \\
        K^{\mathrm{(inter)}}_{\alpha \beta \gamma}
        &=
        -\frac{eg_S \mu_B}{4} \sum_n \int_{\bm{k}} 2f(E_{n\bm{k},\tilde{\bm{q}}} )
        \mathrm{Re}[
        \varepsilon_{\alpha \gamma \delta} 
        \bra{ u_{n\bm{k},\tilde{\bm{q}}} } \tilde{\sigma}_{\delta} \ket{ u_{m\bm{k},\tilde{\bm{q}}} }
        \braket{ u_{m\bm{k},\tilde{\bm{q}}} | \partial_{\tilde{q}_{\beta}/2} u_{n\bm{k},\tilde{\bm{q}}} }
        - (\alpha \leftrightarrow \beta)] \nonumber \\
        &=
        \frac{eg_S \mu_B}{4} \sum_n \int_{\bm{k}} f(E_{n\bm{k},\tilde{\bm{q}}} )
        \varepsilon_{\alpha \gamma \delta}
        \frac{\partial^2 E_{n\bm{k},\tilde{\bm{q}}} }{\partial h_{\delta} \partial (\tilde{q}_{\beta}/2) }
        - (\alpha \leftrightarrow \beta).
\end{align}
\end{subequations}
The intraband term originates from only the Bogoliubov Fermi surface.
Then, summing these two terms, the spatially-dispersive superfluid weight is
\begin{equation}
    K_{\alpha \beta \gamma}
    =
    \frac{e g_S \mu_B}{2} \varepsilon_{\alpha \gamma \delta} \frac{\partial^2 F(\Delta,\tilde{\bm{q}})}{\partial h_{\delta} \partial (\tilde{q}_{\beta}/2 )}
    -
    (\alpha \leftrightarrow \beta).
\end{equation}
This formula is also an approximation neglecting the $\tilde{q}$ dependence of the gap function, as discussed above.

\section{Derivation of the thermodynamic relation (Eq.~\ref{thermo_relation1})}
The COM momentum without a supercurrent $\tilde{\bm{q}}_0$ is determined by the minimizing the free energy density $F[\tilde{\bm{q}},\bm{h},T]$, satisfying $\bm{j}(\tilde{\bm{q}}_0,\bm{h},T) = \partial F[\tilde{\bm{q}}_0,\bm{h},T]/\partial \tilde{\bm{q}}_0 = 0$. When the magnetic field is varied by $\delta \bm{h}$ with the constraint of a vanishing supercurrent, the COM momentum should be changed accordingly. This constraint is given by
\begin{align}
    &j^{\alpha}(\tilde{\bm{q}}_0 + \delta \tilde{\bm{q}}_0,\bm{h}+\delta \bm{h},T) = 0 \nonumber \\
    &=
    (\partial j^{\alpha}/\partial \tilde{q}^{\beta}_0) \delta \tilde{q}^{\beta}_0 + (\partial j^{\alpha}/\partial h^{\beta}) \delta h^{\beta} + \mathcal{O}(\delta^2).
\end{align}
Here, 
$\partial j^{\alpha}/ \partial \tilde{q}^{\beta}_0 = \partial^2 F[\tilde{\bm{q}}_0,\bm{h},T]/\partial \tilde{q}^{\alpha}_0 \partial \tilde{q}^{\beta}_0$ and $\partial j^{\alpha}/ \partial h^{\beta}= \partial^2 F[\tilde{\bm{q}}_0,\bm{h},T]/\partial \tilde{q}^{\alpha}_0 \partial h^{\beta}$ are the superfluid weight and the superconducting Edelstein coefficient, respectively. In the two-dimensional Rashba SOC system used in the main text, when the magnetic field is applied along the $y$-axis ($\bm{h} = h \bm{e}_y$), the COM momentum is along the $x$-axis ($\tilde{\bm{q}}_0 = \tilde{q}_0 \bm{e}_x$).
Then, $\partial j/ \partial \tilde{q}_0 = \partial^2 F[\tilde{q}_0,h,T]/ \partial \tilde{q}^2_0 = K_{xx}/4$ and $\partial j/ \partial h = \partial^2 F[\tilde{q}_0,h,T]/ \partial \tilde{q}_0 \partial h = -K_{zxx}/2$.
Here, we set $e^2=1$ and $-eg_s \mu_B/2 =1$.
Thus we obtain Eq.~\ref{thermo_relation1}. For other helical superconductors, a similar thermodynamic relation can be derived using the above discussion making the measurement of the COM momentum possible for all helical superconductors.

\section{Derivation of the thermodynamic relation (Eq.~\ref{CC_relation})}
\begin{figure*}[t]
\includegraphics[width=0.33\linewidth]{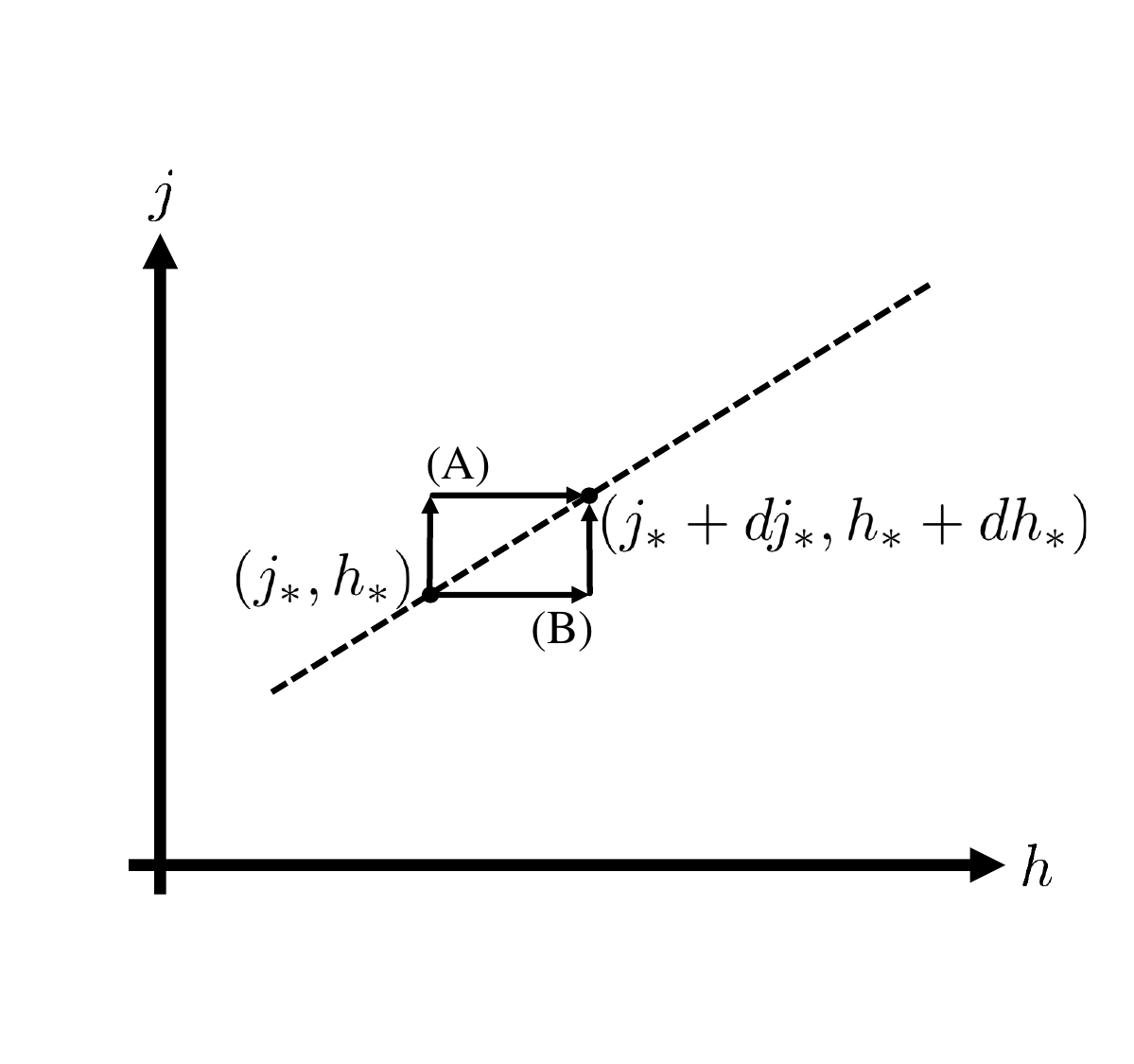}
\includegraphics[width=0.32\linewidth]{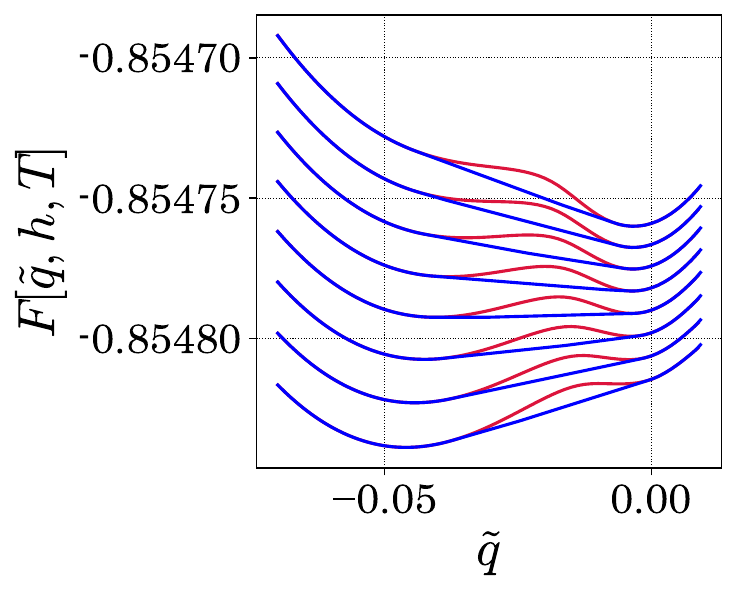}
\includegraphics[width=0.32\linewidth]{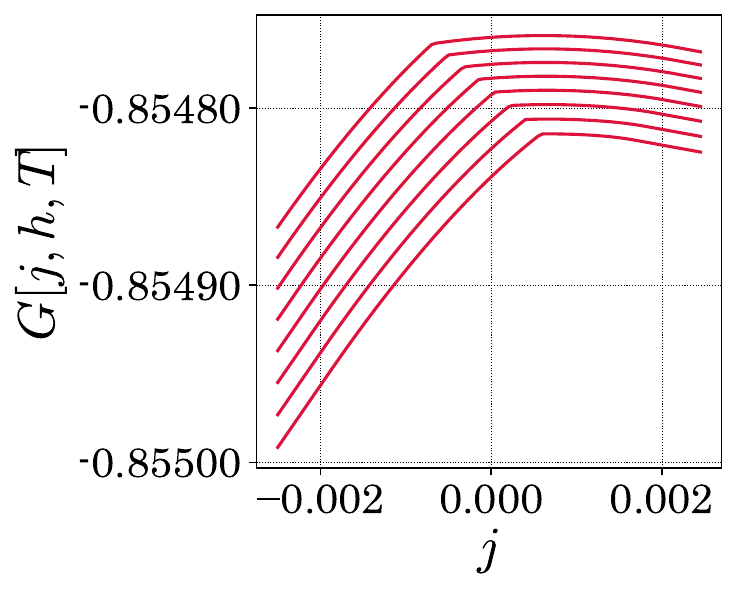}
\caption{(Left) The diagram of the first-order transition line on the $j$-$h$ plane. (Middle) The momentum dependence of the free energy density $F[\tilde{q},h,T]$. The curves for $h=0.057$, 0.058, 0.059, 0.06, 0.061, 0.062, 0.063, and 0.064 are in order from top to bottom. The red curves represent the equation of the free energy density obtained through the mean-field approximation, and the blue curves represent the free energy density transformed by the Legendre transformation twice. (Right) The supercurrent $j$ dependence of $G[j,h,T]$ is plotted. The curves for $h=0.057$, 0.058, 0.059, 0.06, 0.061, 0.062, 0.063, and 0.064 are in order from top to bottom.}  \label{free_energy_fig}
\end{figure*}
The free energy density $F[\tilde{q},h,T]$ must be a continuous and convex function. We introduce $G[j,h,T] = \min_{\tilde{q}}(F[\tilde{q},h,T]) - j\tilde{q})$ by the Legendre transformation. This function is also a continuous and convex function, and $\tilde{q}$ and $j$ have a one-to-one correspondence.
If the system has a first-order transition and momentum jumps, the $G$ function contains a non-differentiable point $j_*$. The difference between the right and left derivatives of this point ( $\partial G[j_* - 0,h,T]/\partial j_* - \partial G[j_* + 0,h,T]/\partial j_*$ ) equals the jump of the momentum.

Let us consider a point $(j_* , h_*)$ in the first-order transition line on the $j$-$h$ plane, as shown by the dashed line in Fig.~\ref{free_energy_fig} (left).
When going to another point $(j_* + dj_* , h_* + dh_*)$ on this line, two paths, (A) and (B), can be considered. First, considering the change of the $G$ function along the path (A), we obtain
\begin{align}
    &G^{\mathrm{(A)}}[j_* +d j_*,h_*+dh_*,T] - G[j_*,h_*,T] \nonumber \\
    &=\frac{\partial G[j_*+0,h_*,T]}{\partial j_*} dj_* + \frac{\partial G[j_*,h_*-0,T]}{\partial h_*} dh_* \nonumber \\
    &= \tilde{q}^{(\mathrm{A})}(j_*,h_*,T) dj_* + m^{\mathrm{(A)}}(j_*,h_*,T) dh_*.
\end{align}
Second, considering the other path (B), the change of the $G$ function reads
\begin{align}
    &G^{\mathrm{(B)}}[j_* +d j_*,h_*+dh_*,T] - G[j_*,h_*,T] \nonumber \\
    &=\frac{\partial G[j_*-0,h_*,T]}{\partial j_*} dj_* + \frac{\partial G[j_*,h_*+0,T]}{\partial h_*} dh_* \nonumber \\
    &= \tilde{q}^{(\mathrm{B})}(j_*,h_*,T) dj_* + m^{\mathrm{(B)}}(j_*,h_*,T) dh_*.
\end{align}
Because the $G$ function is continuous, the two above equations are equal, and we obtain a thermodynamic relation
\begin{align}
    \frac{dj_*}{dh_*} = - \frac{m^{\mathrm{(B)}}(j_*,h_*,T) - m^{\mathrm{(A)}}(j_*,h_*,T)}{\tilde{q}^{(\mathrm{B})}(j_*,h_*,T) - \tilde{q}^{(\mathrm{A})}(j_*,h_*,T)}.
\end{align}
This relation is the extension of the Clausius–Clapeyron equation known in the first-order transition in the liquid-gas transition.

\section{Numerical calculation of Eq.~\ref{CC_relation}}
The free energy density $F[\tilde{q},h,T]$ in our model at $T=10^{-3}$ is plotted in Fig.~\ref{free_energy_fig} (middle). The curve colored by red is calculated by Eq.~(\ref{free_energy}), and the convexity is violated. This is an obstacle caused by the mean-field approximation. Using the Legendre transformation, we plot $G[j,h,T]$, which shows the non-differentiable point $(j_*,h_*)$ at each magnetic field, resulting in the first-order transition line in the $j$-$h$ plane.
Using the Legendre transformation again, the revised free energy density $F$ is plotted by blue curves in Fig.~\ref{free_energy_fig} (middle). This $F$ function restores the convexity.
The region where convexity was violated is connected by a straight line, and its slope corresponds to the value of $j$ that gives a non-differentiable point in the $G$ function.

\end{document}